# Probing top-gated field effect transistor of reduced graphene oxide monolayer made by dielectrophoresis

K. S. Vasu,[1] Biswanath Chakraborty,[1] S. Sampath,[2] and A. K. Sood[1,*]

[1]*Department of Physics, Indian Institute of Science, Bangalore-560012, India.*

[2]*Department of Inorganic and Physical chemistry, Indian Institute of Science, Bangalore-560012, India.*

*Corresponding author: asood@physics.iisc.ernet.in

**Abstract**

We demonstrate top-gated field effect transistor made of reduced graphene oxide (RGO) monolayer (graphene) by dielectrophoresis. Raman spectrum of RGO flakes of typical size of 5μm x 5μm show a single 2D band at 2687 cm$^{-1}$, characteristic of a single layer graphene. The two probe current – voltage measurements of RGO flakes, deposited in between the patterned electrodes with a gap of 2.5 μm using a.c. dielectrophoresis show ohmic behavior with a resistance of ~ 37kΩ. The temperature dependence of the resistance (R) of RGO measured between temperatures 305 K to 393 K yields temperature coefficient of resistance $[dR/dT]/R \sim -9.5 \times 10^{-4}$ K$^{-1}$, same as mechanically exfoliated single layer graphene. The field effect transistor action was obtained by electrochemical top-gating using solid polymer electrolyte (PEO + LiClO$_4$) and Pt wire. Ambipolar nature of graphene flakes is observed upto a doping level of ~ 6 X 10$^{12}$/cm$^2$ and carrier mobility of ~ 50 cm$^2$V$^{-1}$sec$^{-1}$. The source – drain current characteristics shows a tendency of current saturation at high source – drain voltage which is analyzed quantitatively by a diffusive transport model.

## I. Introduction

Single, bilayer and a few layer graphenes continue to be of immense interest not only for their fascinating physical properties but also for their potential device applications [1-7]. However bulk processing of single layer graphene and depositing at desirable locations to fabricate devices and sensors are still challenging issues. Graphene can be prepared by various methods such as micro-mechanical cleavage [2], chemical vapor deposition [8], thermal annealing of SiC [9] and arc discharge method [10]. In recent years chemical methods [11, 12] of preparing graphene with some functional oxide groups, referred to as reduced graphene oxide (RGO), are getting attention to get high yield of single and a few layer graphene flakes which can be deposited on desirable substrates to make large area conducting coatings and to fabricate devices and sensors. The latter can be fabricated in two ways: (i) the RGO dispersion is drop-cast on the desired substrate and electrodes are deposited on the flakes using top-down approach [12], (ii) the graphene oxide (GO) dispersion prepared using Hummer's method or Brodie's method is used to deposit the graphene oxide flakes in between the prefabricated electrodes using dielectrophoresis [13] and flakes are reduced by Hydrazine treatment [14]. The obvious limitation of the first method is that graphene flakes cannot be deposited at predefined locations on the substrate and in the latter method, the GO sheets may not be completely reduced. Recently, single and a few layer graphene sheets made by chemical methods have also been deposited using dielectrophoresis, but these devices did not show any gate response [15, 16].

Here we demonstrate field effect gating of RGO single layer deposited between prefabricated electrodes using dielectrophoresis. This is achieved using solid polymer electrolyte (PEO + LiClO$_4$) and Pt wire [17,18,19,20] in top gate geometry where the Fermi level can be significantly shifted by applying small gate voltages ($\sim$ 1V) due to large gate capacitance ($\sim$ 1.5 µF/cm$^2$). The RGO single layer is characterized by 2D Raman band at 2687 cm$^{-1}$ and temperature coefficient of resistance is $\sim$ -0.095 % C$^{-1}$, same as that of mechanically exfoliated monolayer graphene [21]. We note that the electrolytic gate capacitance is $\sim$ 125 times higher than the gate capacitance of 300nm SiO$_2$ and $\sim$ 3 times higher than the top gate capacitance (550 nF/cm$^2$) using HfO$_2$ as a high K dielectric material [22]. Our RGO FET devices show on – off ratio of $\sim$ 4. The source – drain current ($I_{DS}$) starts deviating from linear dependence on source – drain voltage ($V_{DS}$) at higher $V_{DS}$ as seen recently in FET devices made of mechanically exfoliated single and bilayer graphene [19, 22]. This is understood in diffusive transport model incorporating spatial – dependence of induced carriers between the source and drain electrodes.

## II. Experimental

Single and a few layer RGO dispersions are synthesized using graphitic oxide prepared by modified Hummer's method [23, 24] using exfoliated graphite (Stratmin exfoliated graphite, exfoliated at 800$^0$C termed as SE 800) [25]. Aqueous dispersions of graphitic oxide of concentration $\sim$ 1mg/ml are obtained by sonication for 1 hour, followed by centrifuging at 5000 rpm for 15minutes and vacuum filtration through Whatman filter paper to remove the sediment and large particles. A reduction of this filtrate is carried out by adding 3ml of anhydrous Hydrazine ($N_2H_4$) to 10ml of graphene oxide dispersion in nitrogen filled flask and allowed to stir for 1 day at temperature of $95^0C$ in oil bath. RGO suspension and flakes were characterized using atomic force microscopy (Veeco Nanoscope IV A), UV – Visible spectra (Perkin – Elmer Lamda 35) and Raman spectroscopy (Witec spectrometer with 514.5nm excitation).

Source and drain electrodes are patterned on 300nm $SiO_2$ /Si using photolithography (MicroTech Laser Writer LW 405) followed by RF sputtering of 10 nm Cr and 30 nm Au. The source – drain separation (L) is 2.5 μm and width of the electrodes (W) is 5 μm. A drop of suspension (2 μl) of reduced graphene oxide is then dropped in-between the electrodes on the substrate, and an alternating electric field of 1 MHz and Vpp = 10V is applied between the electrodes followed by keeping the device under an Infrared lamp to dry up the solution. Temperature dependence of the resistance of the RGO flakes was measured from room temperature to $120^0C$ using a hot stage in a cryostat. Top gating was achieved by using solid polymer electrolyte consisting of lithium perchlorate ($LiClO_4$) and poly-ethylene oxide (PEO) in the ratio 0.12: 1. The gate voltage was applied by placing platinum electrode in the polymer layer and FET electrical measurements were carried out at room temperature using Keithely 2400 source meters.

**III. Results and Discussions**

Fig. 1(a) shows UV – Visible absorption spectra of the GO and the RGO suspensions. It can be seen that the absorption peaks at 230nm (5.44eV) for GO and at 280nm (4.42eV) for RGO suspensions are in agreement with earlier reports [11]. The absorption maximum in RGO is associated with the energy gap between the $\Pi - \Pi^*$ bands at the M – symmetry point in Brillouin zone of graphene. Fig.1(b) shows Raman spectrum of a typical RGO flake deposited between the electrodes, displaying G – band at 1594 $cm^{-1}$ ,defect - induced mode D – band at 1349 $cm^{-1}$, 2D band at 2687 $cm^{-1}$, combinational modes D+G band at 2933 $cm^{-1}$ and 2G band at 3191 $cm^{-1}$. It is seen that the G – band is blue shifted as compared to G – band in pristine monolayer and a few layer graphene at ∼ 1582 $cm^{-1}$. The peak position of the 2D band is similar to that of a monolayer graphene prepared using mechanical exfoliation [7].

The inset of Fig. 2 shows I – V data of the graphene flake at zero gate voltage, yielding the resistance of RGO to be 36.6 kΩ. Fig. 2 shows the normalized resistance (R (T)/ R (305 K)) of the RGO flakes deposited between the electrodes as a function of temperature between 305K and 393K. The resistance of the RGO decreases by ∼ 9%, on increasing the temperature to 393 K. The temperature dependence of the resistance of the RGO flake [dR/dT]/R ∼ - 9.5 X $10^{-4}$ /K is in good agreement with temperature dependence of the single layer graphene prepared through mechanical exfoliation and also with the theoretical calculations by Vasko and Ryzhii [26].

Fig.3 shows the device resistance as a function of top-gate voltage, using the source-drain voltage $V_{DS}$ of 50mV. The gate dependence shows ambipolar nature of the RGO channel. The application of top gate voltage $V_{TG}$ creates an electrostatic potential difference Ø between the graphene sheet and gate electrode along with a shift of the Fermi level ($E_F$). Therefore, $eV_{TG} = E_F + (n_{ind}e^2/C_{TG})$. $C_{TG}$ is the geometrical capacitance, which is taken to be 1.5μF/$cm^2$ as reported recently [19] and $n_{ind}$ is the carrier density induced by the top gate voltage $V_{TG}$. The Fermi energy of graphene changes as $E_F(n) = \hbar |v_F| \sqrt{\Pi n_{ind}}$, where $|v_F| = 1.1 \times 10^6$ m/s is the Fermi velocity. Using these values of $|v_F|$ and $C_{TG}$, $n_{ind}$ is calculated as a function of gate voltage and is also shown in Fig. 3 (note that $n_{ind}$ is not a linear function of $V_{TG}$).

In diffusive transport model [7, 19, 22], appropriate for a graphene channel of length 2.5 μm at room temperature, the resistance of the device is written as $R = R_C + R_{Channel}$, where $R_C$ is the total contact resistance at the source and drain contacts, $R_{Channel} = L / (Wne\mu)$ is the resistance of the RGO sample and μ is the mobility of carriers, $n = \sqrt{(n_{ind})^2 + (\delta n)^2}$ [27, 19], δn is the unintentionally generated carrier concentration due to charge puddles [28, 29] and due to residual functional groups. Using these relations, the solid line in Fig.3 is a fitted curve with values of μ = 58 cm$^2$/V.sec, δn = 1.2 × 10$^{12}$ /cm$^2$ and $R_C$ of 9.3kΩ (fit shown in Fig.3). The poor fit of the model with the data at negative values of $V_{TG}$ can arise due to different value of contact resistance on the hole doping side, which has not been taken into account in our simple model. We also estimated mobility of the device using $(dI_{DS}/dV_{TG})_{max} = (W\mu C_{TG}V_{DS})/L$. The value of $(dI_{DS}/dV_{TG})_{max}$ = 2.9 × 10$^{-6}$ S which gives μ = 22 cm$^2$/V.sec.

We now discuss another feature of FET characteristics, when drain - source voltages is high. Fig. 4(a) shows $I_{DS}$ as a function of drain – source voltage ($V_{DS}$) for different top gate voltages $V_{TG}$-$V_D$, where $V_D$ is the voltage at Dirac point (maximum resistance). The noticeable feature is that, current -voltage characteristics deviate considerably from linearity, as also seen in mechanically exfoliated single layer graphene [22] and bilayer graphene [19]. The non – linearity in current - voltage curves has been explained earlier by diffusive transport model incorporating the space dependent carrier concentration in graphene channel between the source and drain electrodes. We will use a similar model to see if it explains our present data.

In diffusive transport model, the current $I_{DS}$ can be written as [19, 22]

$$I_{DS} = \frac{W}{L} \int_0^L e n(x) v_d(x) dx \quad (1)$$

where n(x) is the carrier concentration which depends on the position x along the channel and is given by $n(x) = \sqrt{(n_{ind}(x))^2 + (\delta n)^2}$, $v_d$ is the drift velocity given by $v_d(x) = \mu E(x) = \mu \frac{dV(x)}{dx}$, E(x) is the longitudinal electric field and V(x) is the potential drop at any point in the channel. Using the gradual channel approximation Eq. (1) can be written as

$$I_{DS} = \frac{W}{L} e\mu \int_{V_1}^{V_2} \sqrt{(n_{ind})^2 + (\delta n)^2} dV \quad (2)$$

where $V_1 = I_{DS}R_C/2$, $V_2 = V_{DS} - (I_{DS}R_C)/2$ are the potential drops at the source and drain ends respectively and $R_C$ is the total contact resistance. For the FET made of monolayer graphene prepared by mechanical exfoliation, it was seen that velocity of the carriers can show saturation and hence $v_d = \mu E / (1 + (\mu E / v_{sat}))$. In bilayer FET, velocity saturation of the carriers was not needed to fit the $I_{DS} - V_{DS}$ characteristics. In the present case with the RGO device, the mobility of the carriers is much lower as compared to mechanically exfoliated graphene samples and hence velocity saturation is not needed to fit the $I_{DS} - V_{DS}$ characteristics.

Using the above values of mobility $\mu = 58$ cm$^2$/V.sec, the minimum carrier concentration $\delta n = 1.2 \times 10^{12}$ /cm$^2$, and $R_C = 9.3$ k$\Omega$, equation (2) is solved self – consistently to get $I_{DS}$ values for different values of $V_{DS}$. Fig. 4(b) shows the experimental curves along with the theoretical plots for different top gate voltages. The theoretical and experimental curves are in good agreement. From this model, we can also extract $n_{ind}(x)$.

Fig. 5 shows the variation of induced carrier concentration along the channel length of the device at 0.4 V top gate voltage ($V_{TG}$). For $V_{DS} = 0.1$ (i.e. $V_{DS} < V_{TG}$), the induced carrier concentration (electrons) decreases linearly along the channel length from source end to drain end and reaches constant value. When $V_{DS} = V_{TG}$, the carrier concentration at drain end reaches to zero. This is called pinch off region. For $V_{DS} = 1$ V (i.e. $V_{DS} > V_{TG}$), the pinch off point moves towards the source end and after the pinch off point the channel is hole doped upto the drain end.

**IV. Summary**

In conclusion, we have demonstrated FET characteristics of reduced graphene oxide monolayer using top gating arrangement. The temperature coefficient of resistance of the RGO is ⊔ -0.095 % C$^{-1}$ is observed, same as that of mechanically exfoliated monolayer graphene. Nonlinear $I_{DS} - V_{DS}$ characteristics at different top - gate voltages are observed at large values of $V_{DS}$ which could be explained in a diffusive transport model incorporating spatial dependence of induced carrier concentration. Our results show that FET can be fabricated in a very simple manner in large quantities by dielectrophoresis of the chemically prepared reduced graphene oxide suspension, also has potential reality in making RGO based sensors. This kind of FET based sensors opens up the possibility of using RGO with controlled charge densities prepared by chemical methods.

**Acknowledgements**
We thank Department of Science and Technology for financial assistance under the Nanoscience Initiative.

**Figure Captions**

Fig.1 (a) UV absorption spectra of graphene oxide dispersion (thick line) and reduced graphene oxide dispersion (thin line). (b)Raman spectrum of reduced graphene oxide flake showing D, G, 2D, D+G and 2G modes.

Fig. 2: Temperature dependence of the resistance of RGO flake.

Fig.3: Resistance (closed circles) with fitted data (solid line) and Induced carrier concentration (dotted line) as a function of top gate voltage ($V_{TG}$).

Fig.4: (a) Source – drain current as a function of source – drain voltage ($V_{DS}$) for different top gate voltages.(b) Comparison of the experimental data (solid lines) and with the diffusive transport model [Eq. 2] (dashed lines).

Fig. 5: Variation of induced carrier concentration along the channel length at constant gate and source – drain voltage.

Figure.1:

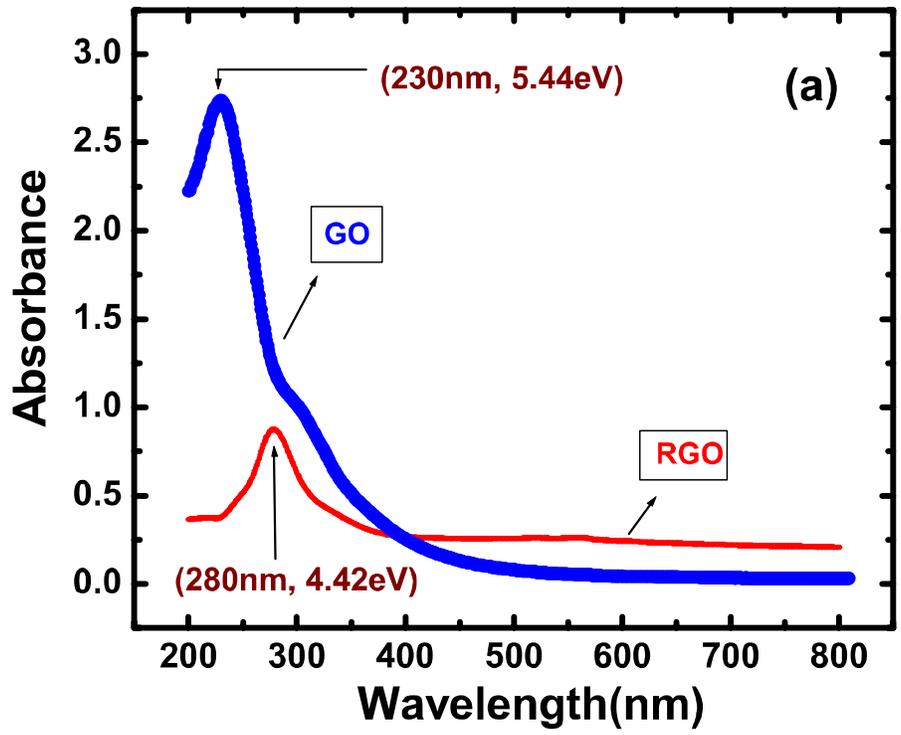

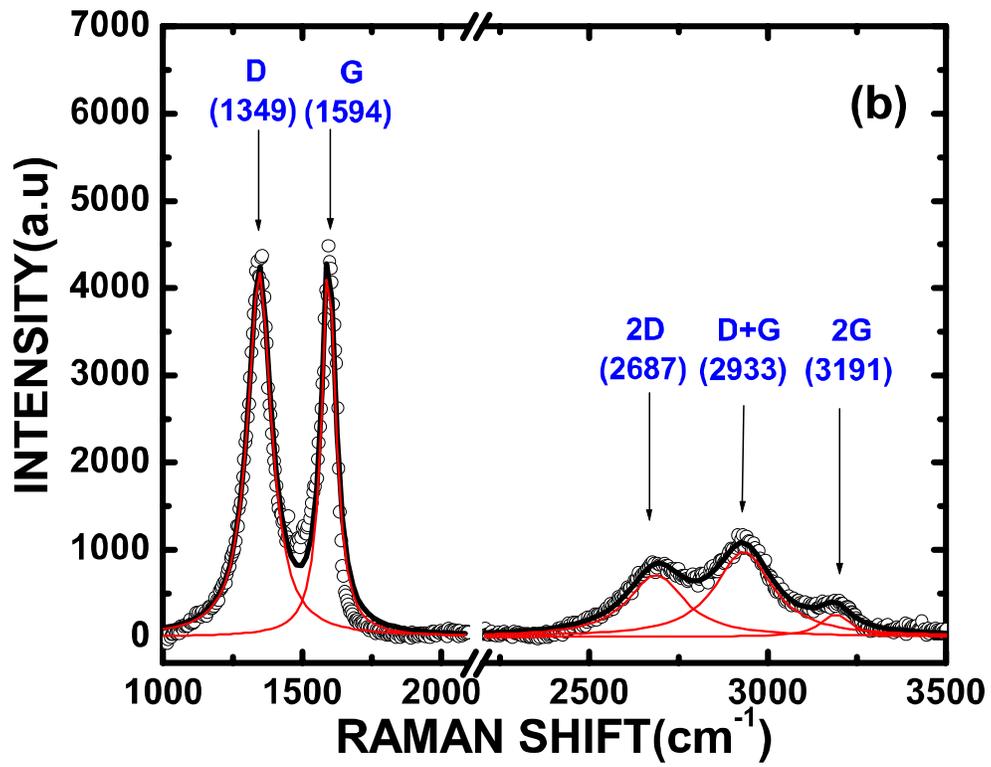

Figure.2:

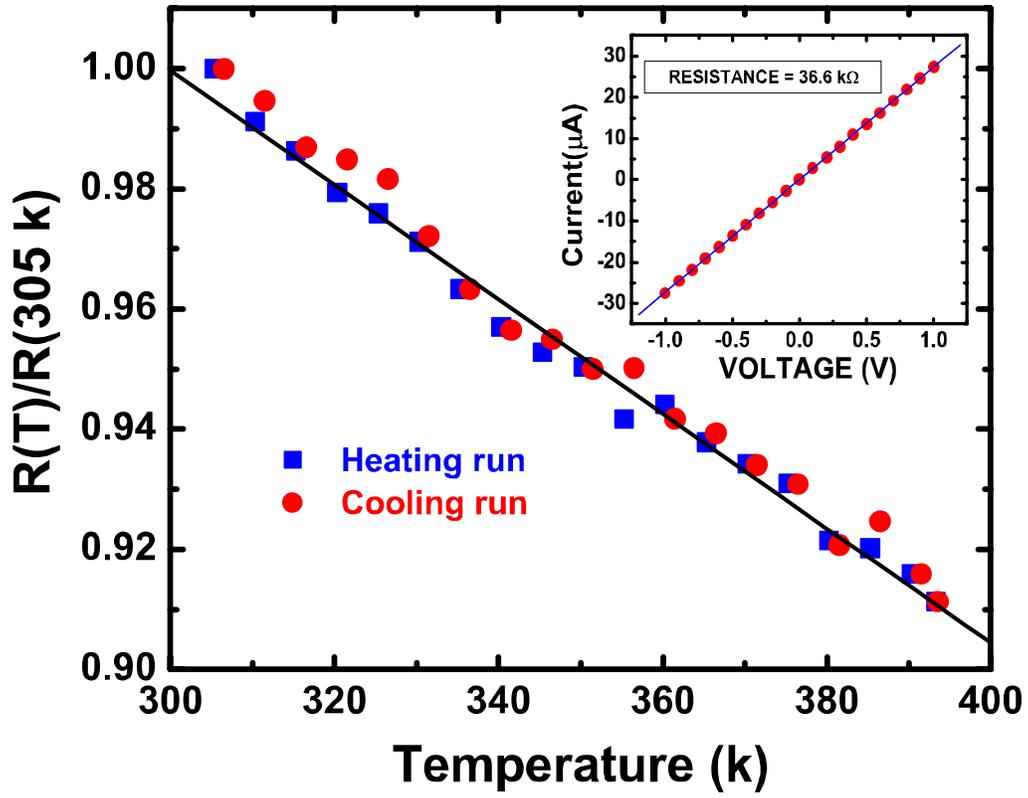

Figure.3:

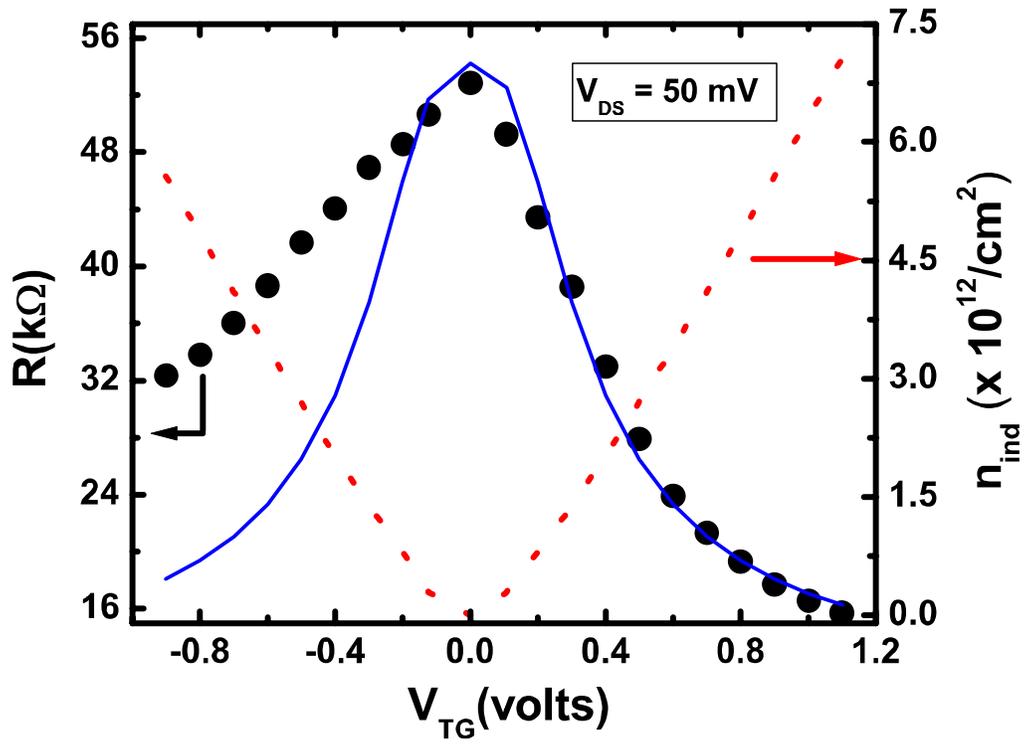

Figure.4:

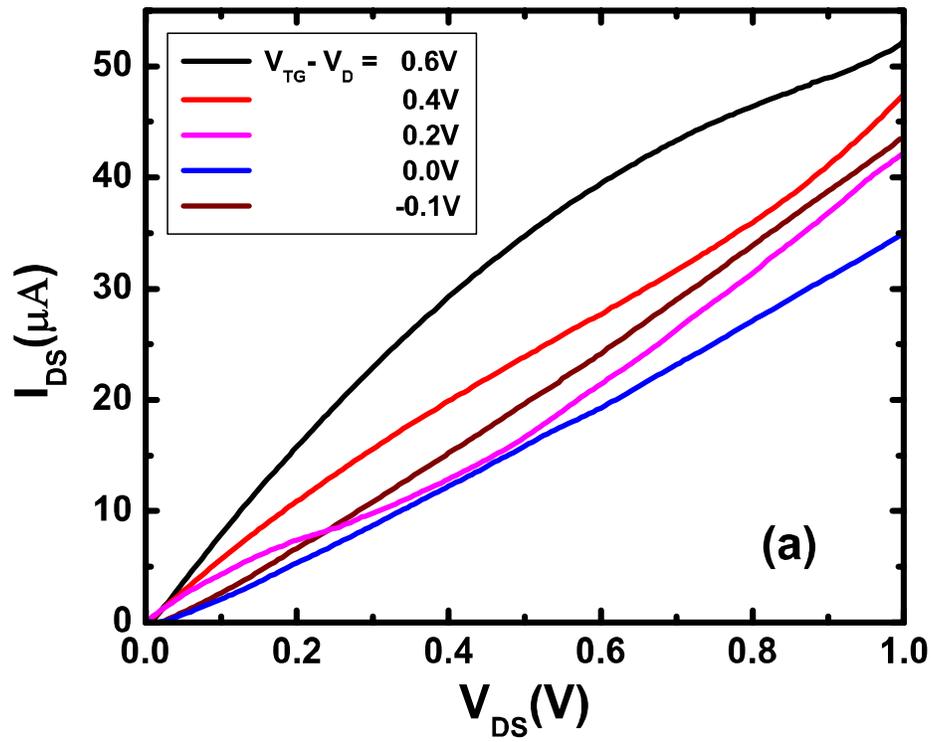

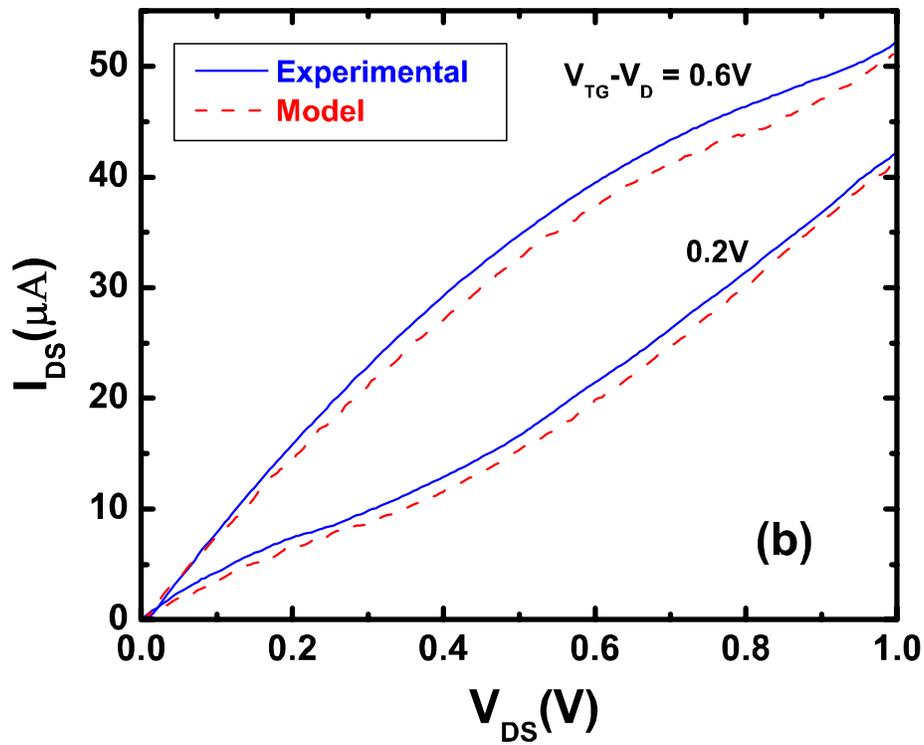

Figure.5:

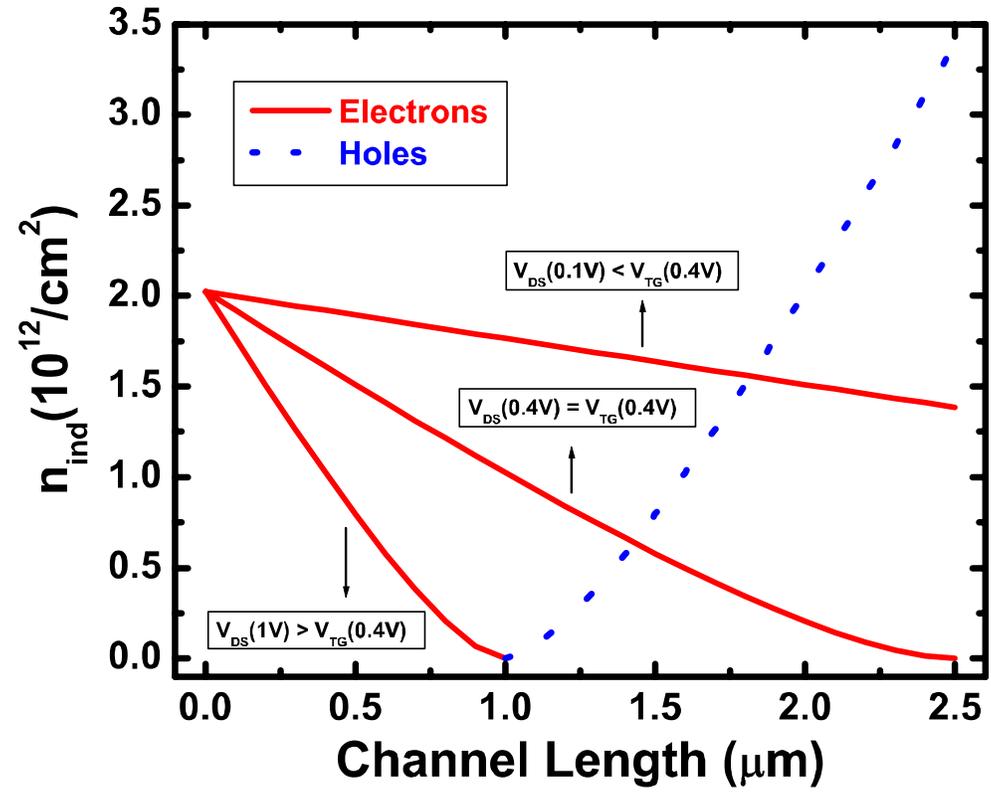